\documentclass[twocolumn,floatfix,prl,aps,showpacs]{revtex4-1}
\usepackage{graphicx,color,amsmath,amssymb}
\usepackage{bm,hyperref}
\newcommand{\PRL}{Phys.\ Rev.\ Lett.}
\newcommand{\PRB}{Phys.\ Rev.\ B}

\newcommand{\be}{\begin{equation}}
\newcommand{\ee}{\end{equation}}

\begin{document}
\title{Anomalously low magnetoroton energies of the unconventional fractional quantum Hall states of composite fermions}
\author{Sutirtha Mukherjee and Sudhansu S. Mandal}
\affiliation{Department of Theoretical Physics, Indian Association for the Cultivation of Science, Kolkata 700032, India}
\date{\today}
\begin{abstract} 

We show a generic formation of the primary magnetorotons in the collective modes of the observed ``unconventional'' fractional quantum Hall effect (FQHE)
states of the composite fermions
at the filling factors 4/11, 4/13, 5/13, 5/17, and 3/8 at very low wavevectors with {\em anomalously} low energies 
which do not have any analogue to the conventional fractional quantum Hall states.
Rather slow decay of the oscillations of the pair-correlation functions in these states are responsible for the low-energy magnetorotons. 
This is a manifestation of the distinct topology predicted before for these FQHE states. 
Experimental consequences of our theory are also discussed.
\end{abstract}.

\maketitle

The two-dimensional interacting electrons in the presence of strong transverse magnetic field create the correlated ground states of the fractional quantum
Hall effect \cite{Tsui82} (FQHE) which are characterized by the filling factors and topology \cite{Laughlin83} when the electrons are spin-polarized. 
The distinction between different topological natures of the FQHE reveals
through the braid statistics \cite{Moore91,Wilczek91,Nayak96,Read00} of the quasiparticles, through the structure of the gapless edge modes \cite{Wen92}, 
and through the nature of the bulk collective modes \cite{Girvin85,Dev92,Scarola00,Dwipesh09}.
A fair amount of interest has now been rejuvenated \cite{Bellani10,Pan14,Arnold14} for the study of certain ``unconventional'' FQHE at the 
filling factors $\nu = 4/11$, 5/13, and 3/8 which
are in between two conventional FQHE filling factors 1/3 and 2/5. While the conventional FQHE states correspond to the integer quantum Hall effect of 
the composite fermions \cite{Jain89} (CFs) which are quasiparticles consist of an electron and even number $(2p)$ of quantized vortices denoted as $^{2p}$CFs, 
the unconventional FQHE states are formed due
to the nontrivial correlations \cite{Arek04,Mukherjee12,Mukherjee13} of the CFs in the partially filled $\Lambda$ levels ($\Lambda$Ls)--effective 
Landau-like levels of the CFs. Moreover, the topology of
the FQHE states of these CFs is distinct from that for the electrons at the same filling factor.
In this paper, we calculate spinless collective modes of these FQHE states and show that their unconventional topologies reflect in producing
{\em anomalously} low magnetoroton energies.

The origins of some of the unconventional incompressible FQHE states in a range of filling factors between two prominent Jain states \cite{Jain89} are
as follows.
In the range $2/5 > \nu > 1/3$ ($1/3 > \nu > 2/7$), the composite fermion filling factor $\nu^* = 1 + \bar{\nu}$ of $^2$CFs ($^4$CFs) is related
with electron filling factor as $\nu = \nu^*/(2\nu^*+1)$ ($\nu = \nu^*/(4\nu^*-1)$), where $0 <\bar{\nu} <1$. 
We assume all the electrons are spin-polarized.   
The $^2$CFs with $\bar{\nu}=1/3, \,2/3,$ and $1/2$ respectively constitute $\nu = 4/11,\, 5/13$, 
and $3/8$.  The partially filled second $\Lambda$L with filling factors $\bar{\nu} = 1/3$, and its particle-hole conjugate
2/3  are characterized by W\'ojs, Yi, and Quinn (WYQ) state \cite{Arek04} which is the ground state of Haldane pseudo-potential \cite{Haldane83} 
$V_3$ that minimizes occupation
with relative angular momentum $3$ between any two particles.
The even-denominator $3/8$ state is characterized \cite{Mukherjee12} by the anti-Pfaffian pairing correlation \cite{Moore91,Levin07,Fisher07} 
in the half-filled second $\Lambda$L that makes it a nonabelian FQHE state. 
The states of $^4$CFs at $\nu= 4/13,\, 5/17$, and $3/10$ have identical character \cite{Mukherjee15} with the states of $^2$CFs at $\nu =4/11,\,5/13$, and $3/8$ respectively.

We calculate the dispersion of Girvin-MacDonald-Platzman (GMP) mode \cite{Girvin85}  for the
filling factors $\nu=4/11,\, 5/13$, and 3/8 in the range $2/5 > \nu > 1/3$, and for $\nu = 4/13$ and $5/17,$  in the range $1/3 > \nu > 2/7$
which are some of the unconventional FQHE states observed \cite{Pan03,Smet03,Bellani10,Pan14,Arnold14} and have recently been 
proved \cite{Mukherjee12,Mukherjee13,Mukherjee15} as incompressible states. 
We show two prominent, apart from several weaker, magnetorotons for all these states: the position of the secondary one in  
either side of $\nu =1/3$ matches with that for the nearest prominent states at $\nu = 2/5$ or $2/7$; the primary one occurs at a very low 
wavevector and more importantly its energy is extremely low compared to that for the Jain states. 
These are the manifestations of unconventional topology of these states, whereas the conventional topology at $\nu=4/11$ produces just one magnetoroton \cite{Goerbig04}
and whose energy is much higher. We find that slow decay of the oscillations in the pair-correlation
function, $g(r)$, around its long-range limiting value causes a magnetoroton having very low energy.

We employ the single mode approximation (SMA) \cite{Girvin85}  to obtain the dispersion of the GMP collective modes of the FQHE states with unconventional topology.
In the SMA, the spin-conserving excited state may be obtained by operating lowest Landau level (LLL)-projected number-density 
operator on the ground state at a given filling factor leading to the expression for the dispersion \cite{Girvin85}  of the excitation energies:
\begin{eqnarray}
 \Delta(k) &=& 2 [\bar{S}(k)]^{-1}
              \int \frac{d\bm{q}}{(2\pi)^2} \sin^2\left(\frac{\bm{k}\times \bm{q}}{2} l^2\right) e^{-k^2 l^2/2} \nonumber \\
           &\times&     \left[ v(\vert\bm{q}-\bm{k}\vert)e^{\bm{k}\cdot (\bm{q}-\bm{k}/2) l^2} - v(q)\right]  
                \bar{S}(q)
                \label{Dispersion}
\end{eqnarray}
with momenta $\bm{k}$ and $\bm{q}$, Fourier transform of the Coulomb potential $v(q) = 2\pi e^2/(\epsilon q)$. Here $\bar{S}(k) = S(k)-1+e^{-k^2l^2/2}$
is the LLL projection of the static structure factor $S(k)$. We thus determine $S(k)$ below for calculating the GMP mode.

We begin with the calculation of pair-correlation functions $g(r)$ between electrons 
using the ground state wave function for a certain filling factor in the spherical geometry \cite{Haldane83}, 
in which $N$ electrons move on the surface of a
sphere with radius $R =\sqrt{Q} l$, exposed to a magnetic flux $2Q$ in the unit of a flux quantum $\phi_0 = hc/e$ produced by 
a magnetic monopole placed at the center of the sphere. The arc distance between two particles has been considered as $r$. 
The relation between the total (integer) flux, $2Q$, and the total number of electrons, $N$, for all the spin-polarized FQHE states in the 
range $2/5 > \nu > 2/7$ as follows:
\begin{equation}
 2Q = \nu^{-1}(N -1)  - (3-\nu^{-1})(\lambda +2)
 \label{Flux_relation}
\end{equation}
where the so-called ``flux-shift'' $\lambda$ for the CFs in the partially filled $\Lambda$L
is defined in terms of the magnitude of the effective flux as $2\vert Q^*\vert  = \bar{\nu}^{-1}\bar{N}-(\lambda +2)$ with $\bar{N} = N -(2\vert Q^* \vert +1)$ being the number of CFs in the partially filled $\Lambda$L,
and $(2\vert Q^*\vert +1)$ is the degeneracy of the lowest $\Lambda$L.
Recall \cite{Arek04,Mukherjee12,Mukherjee13,Mukherjee15} that the choice of $\lambda$ characterizes the nature of 
correlation in the partially filled $\Lambda$L: $\lambda = 7$ for $\nu = 4/11$ and $4/13$;  $\lambda = -2$ for
$\nu = 5/13$ and $5/17$; and $\lambda = -1$ for 3/8.
The composite-fermion-diagonalization (CFD) method \cite{Mandal02} which is almost exact 
\cite{Mukherjee12,Mukherjee13,Toke14} for finite number of particles has been used to show that the ground states at the filling
factors with these particular $\lambda$ are incompressible. 
Excepting 3/8 state for which we consider the CFD ground state, we have considered the proposed \cite{Mukherjee13,Mukherjee15} CF-WYQ wavefunctions, {\it i.e.},
\begin{eqnarray}
 \Psi_1 &=& P_{\text{LLL}} \prod_{j<k}(u_jv_k-v_ju_k)^{2}\Phi^{\text{WYQ}}_{1+\bar{\nu}}  \label{Eq:Psi1}\\
 \Psi_2 &=& P_{\text{LLL}} \prod_{j<k}(u_jv_k-v_ju_k)^{4}\Phi^{\text{WYQ}}_{-(1+\bar{\nu})} \label{Eq:Psi2}
\end{eqnarray}
respectively for the states corresponding to $^2$CFs and $^4$CFs,
for calculating $g(r)$ as they have significantly high overlap with the CFD ground state, which allows us to calculate for higher number of particles. 
(We have checked that the pair-correlation
calculated using the CFD ground states for smaller number of particles agrees well with that calculated using the wavefunctions up to a maximum accessible distance.) Here spherical spinor variables $u=\cos (\theta/2)e^{-i\phi/2}$ and $v= \sin (\theta/2) e^{i\phi/2}$, $P_{\text{LLL}}$ denotes projection into the LLL,
and $\Phi^{\text{WYQ}}_{\pm(1+\bar{\nu})}$ denotes wavefunction at filling factor $1+\bar{\nu}$ when the partially filled
second Landau level would have WYQ correlation \cite{Arek04}, with $\pm$ referring to the sign of effective magnetic flux, $2Q^*$.

\begin{figure}
\includegraphics[width=8cm]{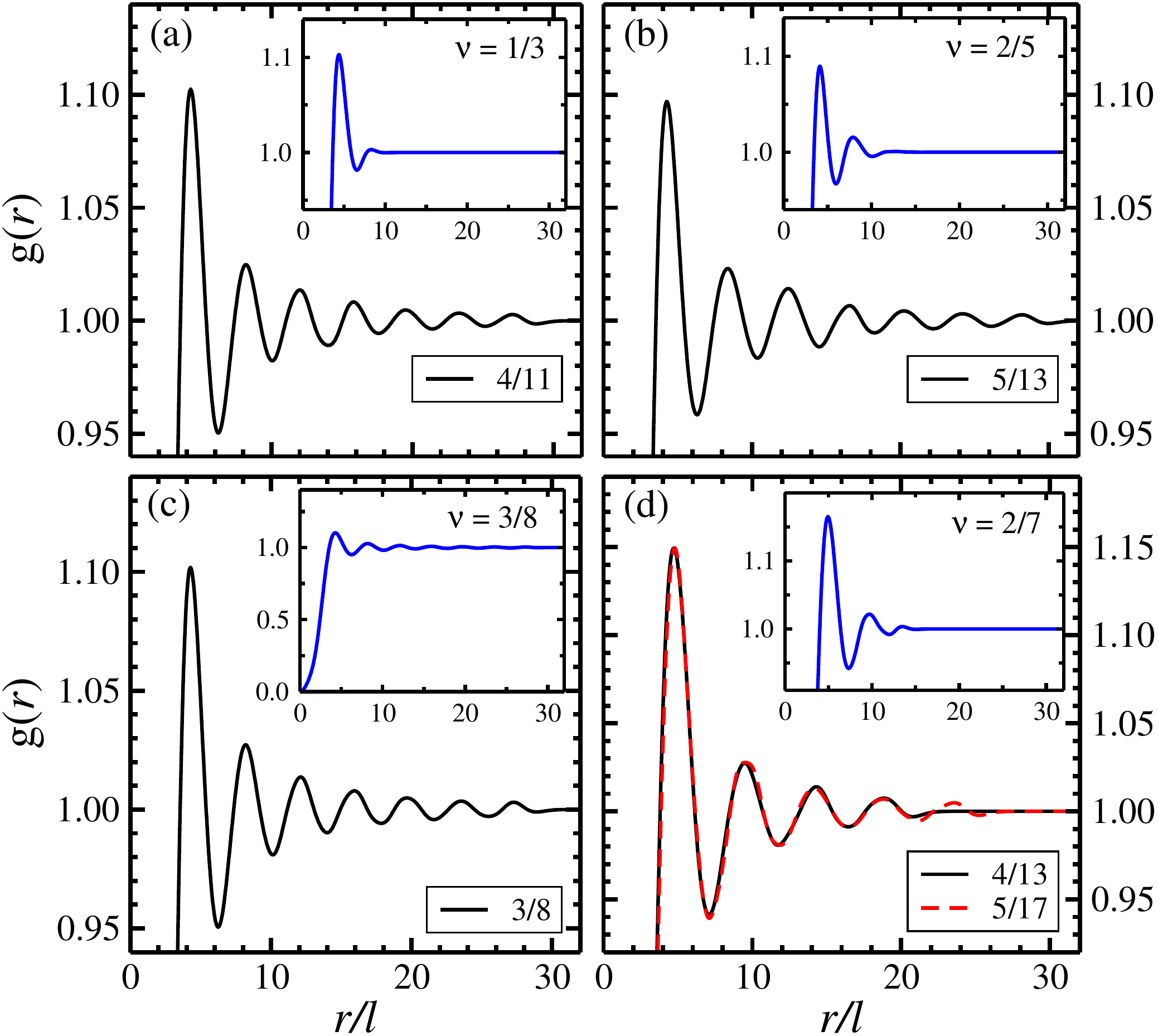}
\caption{(Color online) Calculated Pair-correlation functions $g(r)$ for finite systems and their thermodynamic extrapolation in the damped-oscillatory 
form $g(r) = 1+ A (r/l)^{-\alpha} \sin (\beta r/l - \gamma)$ used earlier \cite{KJG97}, with $A,\,\alpha$, $\beta$, and $\gamma$ being numerical constants,
for (a) $\nu=4/11$, (b) $\nu=5/13$, (c)$\nu =3/8$, and (d) $\nu=4/13$ and 5/17. 
All of these are zoomed for showing oscillations in $g(r)$. Inset (c) shows $g(r)$ for $\nu=3/8$ without zooming. 
Insets of (a), (b), and (d) respectively show $g(r)$ for $\nu=1/3$, 2/5, and 2/7.}
\label{Fig:Correlation}
\end{figure}

We employ the Monte Carlo method with Metropolis algorithm to calculate $g(r)$ for $\nu=4/11$ and 5/13 using the state $\Psi_1$  
for $N=32$ and 36 respectively, $\nu =4/13$ and $5/17$ using the state $\Psi_2$ for $N=24$ and $26$ respectively, and $\nu = 3/8$
using the CFD ground state for $N=24$. We are restricted to consider more number of particles because of determining huge number 
(equal to the number of basis states in a system with particle $\bar{N}$ and flux $2\vert Q^*\vert +2$) of $N\times N$ determinants in each step of the Monte
Carlo and also the projection into the LLL of the states involving $^4$CFs with negative effective flux.
We find that $g(r)$ oscillates around its long range value $1.0$ with smaller decay rate in comparison to the neighboring Jain states, 
but the amplitude of the oscillation in the calculated data for finite systems 
does not die out completely to obtain required $g(r)=1$
at large distances. We therefore extrapolate the numerical data up to large distances using the damped-oscillatory form
$g(r) = 1+A(r/l)^{-\alpha}\sin (\beta r/l-\gamma)$ used earlier \cite{KJG97}, where numerical constants $A,\,\alpha,\,\beta,\,$ and $\gamma$ which are different
for different filling factors, are determined by fitting the available numerical data. 
We thus obtain more oscillations in 
$g(r)$ before it converges to its large distance limit. For example, in the case of $\nu = 4/11$, there are three oscillations \cite{Mukherjee13} obtained using the CFD 
ground state for
$N=28$ and four oscillations using the wavefunction $\Psi_1$ up to $N=32$ for the finite systems, but our thermodynamically
extrapolated $g(r)$ acquires three more oscillations before it converges to unity. 
Figure~\ref{Fig:Correlation} shows $g(r)$ for the thermodynamic systems at different filling factors in the range $2/5 \geq \nu \geq 2/7$. While $g(r)$ for $\nu =1/3$ has 
one maximum, $\nu =2/5$ and 2/7 have two maxima each, and for any state in Jain sequence \cite{Jain89} with $\nu = n/(2n \pm 1)$ has $n$ maxima \cite{Park00}, 
$g(r)$ for the unconventional states have several 
maxima and the number of maxima does not match, in general, with the numerator of the filling factor. 
This has direct consequence on the static structure factor and hence on the energy dispersion of 
the collective modes. Recall \cite{Park00} that the energy of the primary roton at $\nu = 5/11$ is much less than the same at $\nu = 2/5$ because the number of
oscillations in $g(r)$ for the former is more than the latter.

The static structure factors may then be calculated using the relation
$S(k) = 1+n_0\int d\bm{r}e^{i\bm{k}\cdot \bm{r}}[g(r)-1]$, where mean electron density  $n_0 = \nu/(2\pi l^2)$. 
As suggested by GMP \cite{Girvin85}, an appropriate form of 
pair-correlation function of a quantum liquid in a FQHE state at the LLL is given by
\begin{equation}
 g(r) = 1-e^{-r^2/2 l^2}+ \sum_m' \frac{2}{m!}\left(\frac{r^2}{4 l^2}\right)^m c_m \,e^{-r^2/4 l^2}
 \label{pair-correlation}
\end{equation}
where prime indicates summation over odd $m$ only. The numerically calculated $g(r)$ and its thermodynamic extrapolation data are fitted with
the functional form (\ref{pair-correlation}) and the upper cut-off value of $m$ is taken to be very large for picking up oscillations in $g(r)$. 
The coefficients $c_m$'s are constrained with
the charge neutrality, perfect screening, and the compressibility sum rules \cite{Girvin85,Hansen82,Girvin84,Rasolt86} when an analogy with the 
two-dimensional one component plasma is invoked, can be expressed in terms of the respective moments of the pair-correlation 
functions: $M_0=-1$, $M_1=-1$, and $M_2 = 2(\nu^{-1}-2) $ with $M_n = n_0\int d\bm{r} (r^2/2)^n [g(r)-1]$. These sum rules ensure that the projected structure factor into
the LLL behaves as $\bar{S}(k) \to (kl)^4(1-\nu)/8\nu$ as $k \to 0$.

\begin{figure}
\includegraphics[width=8cm]{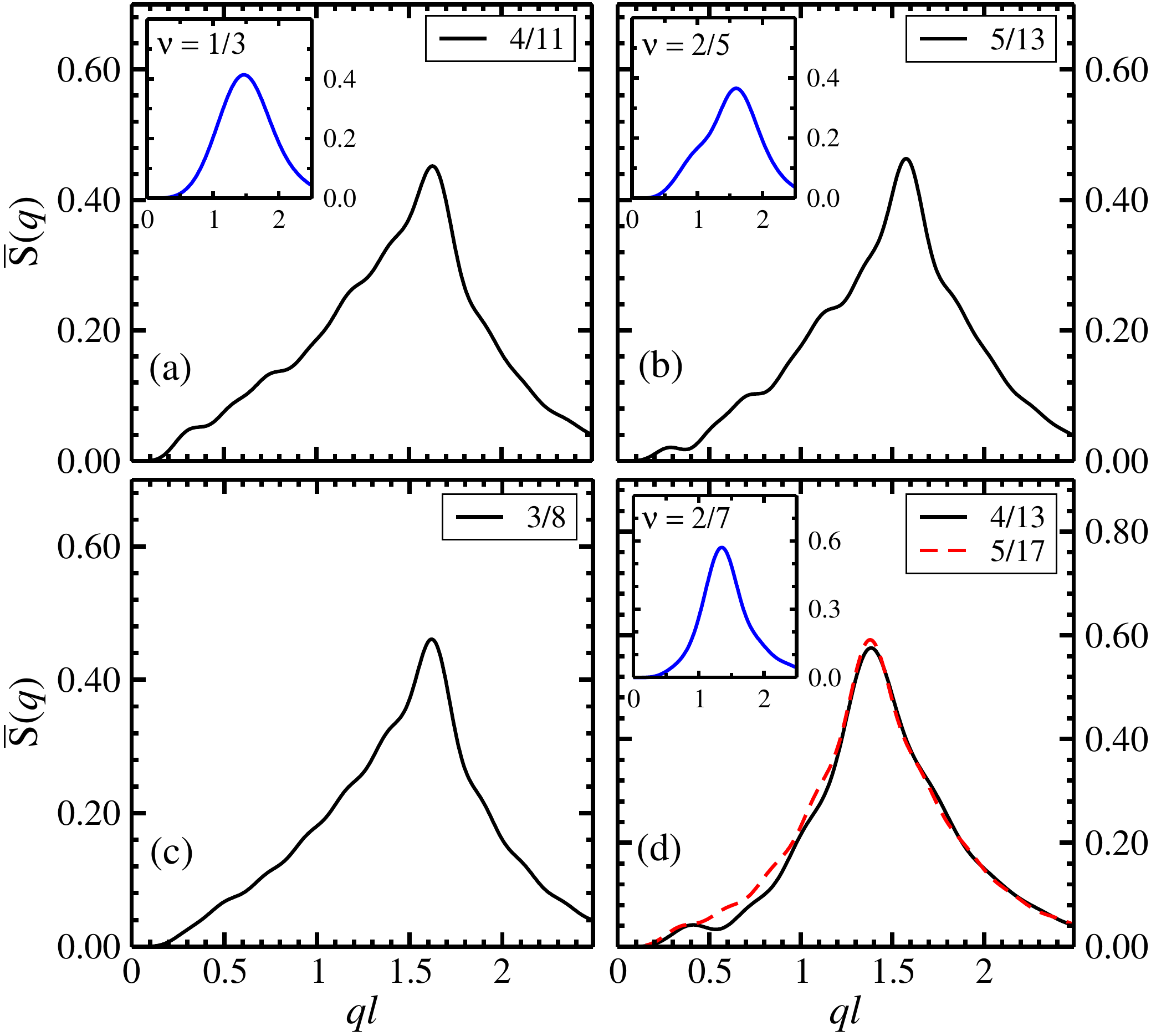}
\caption{(Color online) LLL-projected static structure factor $\bar{S}(k)$ for (a) $\nu =4/11$, (b) $\nu =5/13$, (c) $\nu =3/8$, and (d)
$\nu = 4/13$, and 5/17. $\bar{S}(k)$ for neighboring Jain states at $\nu =1/3$, 2/5, and 2/7 are shown in the insets of (a), (b), and (d) respectively.}
\label{Fig:SFactor}
\end{figure}

We calculate $\bar{S}(k)$ for all the states considered here and show in Fig.~\ref{Fig:SFactor}. $\bar{S}(k)$ for $\nu =1/3$, 2/5, and 2/7 
have also been recalculated \cite{Note1} for comparison and are shown as insets in Fig.~\ref{Fig:SFactor}. 
Owing to the many more number of oscillations in $g(r)$, change in the sign of the slope of $\bar{S}(k)$ occurs several times.
Also, the spectral weights at low through moderate $k$, $(k \lesssim 1.5 l^{-1})$, are more than that of the neighboring Jain states.

\begin{figure}
\includegraphics[width=8cm]{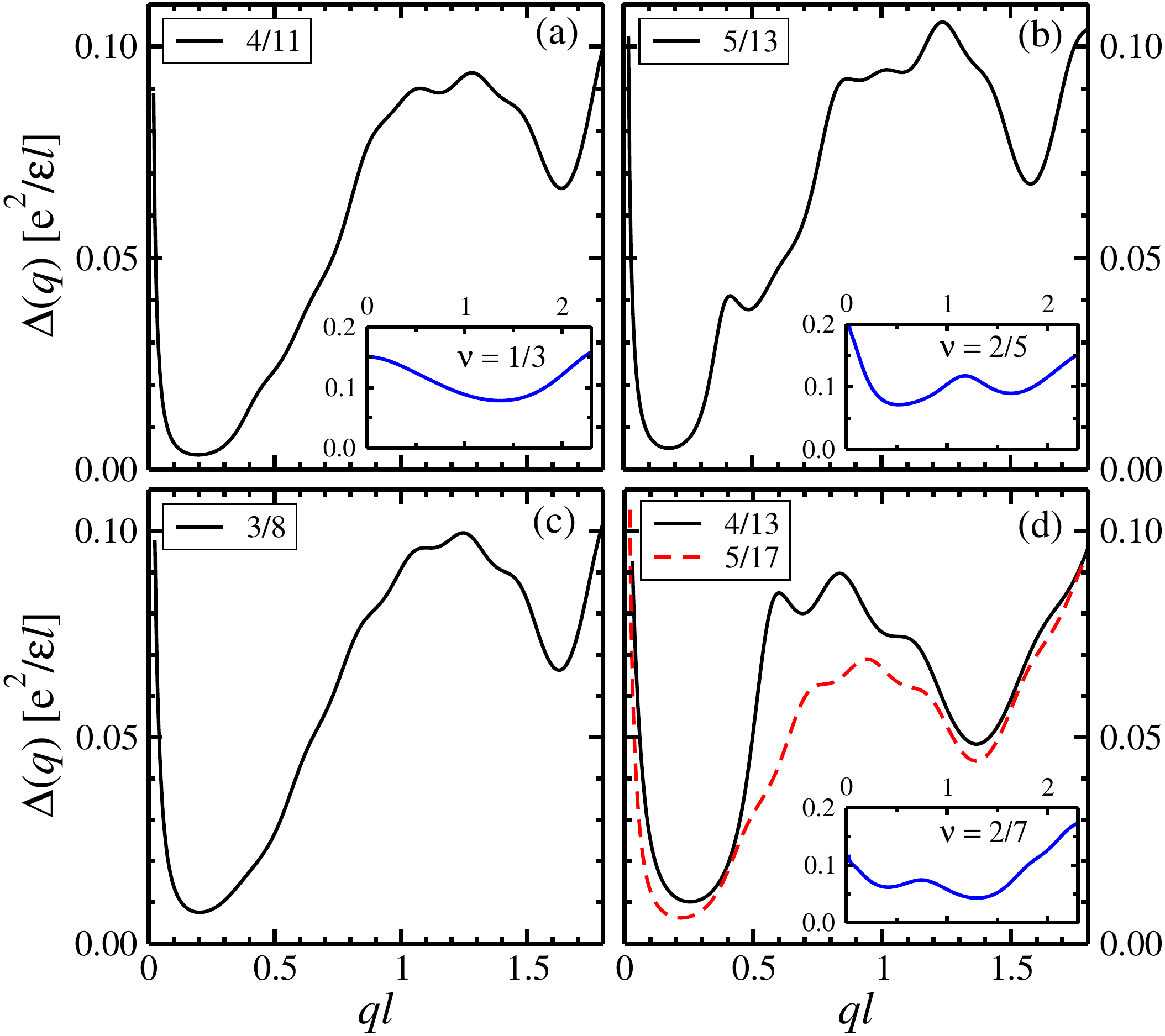}
\caption{(Color online) Dispersion of the GMP mode $\Delta (k)$ in the unit of $e^2/\epsilon l$ for (a) $\nu =4/11$, (b) $\nu=5/13$, (c) $\nu =3/8$, and (d)
$\nu =4/13$ and 5/17. $\Delta (k)$ for neighboring Jain states at $\nu =1/3$, 2/5, and 2/7 are shown as the insets of (a), (b), and (d) respectively.
While there are only one magnetoroton mode at $\nu =1/3$ and two magnetoroton modes at $\nu =2/5$ and $2/7$, all the other states in the range
$2/5 > \nu >2/7$ have two prominent magnetorotons and some not so prominent magnetorotons.}
\label{Fig:Dispersion}
\end{figure}

We next calculate the energy dispersion $\Delta (k)$ of the GMP modes for the unconventional FQHE states at the 
filling factors $\nu = 4/11$, 5/13, 3/8, 4/13, and 5/17 and
show in Fig.~\ref{Fig:Dispersion}. We also show the dispersion of the GMP modes for neighboring Jain states, {\it viz}, $\nu =1/3$, 2/5, and 2/7 
as insets of Fig.~\ref{Fig:Dispersion}. While $\nu =1/3$ has one magnetoroton at $k\simeq 1.5 \,l^{-1}$, $\nu = 2/5$ and 2/7 have two magnetorotons each, 
of which the primary roton minimum occurs at $k\simeq 0.4 \,l^{-1}$ and the secondary roton 
minimum occurs at $k\simeq 1.6 \, l^{-1}$ and $1.4 \, l^{-1}$ respectively. The unconventional FQHE states
in the ranges $2/5 >\nu > 1/3$ and $1/3 >\nu > 2/7$ have two prominent magnetorotons each, of which the position of the 
secondary one matches with that for $\nu= 2/5$ and 2/7 in the respective ranges and the primary one forms at $k \simeq 0.2$--$0.3 \, l^{-1}$.
Several other weaker magnetorotons also form for these states due to the appearance of several changes in the sign of the slopes of $\bar{S}(k)$
at low through moderate $k$, caused by the several oscillations in $g(r)$.
The energy of the primary roton is generically very small and it lies in the range $0.004$--$0.011$ $e^2/\epsilon l$.
As the decay rate of the amplitude of the oscillation in $g(r)$ is less for the unconventional FQHE states compared to
the neighboring Jain states, and the corresponding $\bar{S}(k)$ has higher spectral weight at moderate $k$, the energy of the primary roton becomes
extremely small.

Apart from ignoring ubiquitous disorder and Landau level mixing, we have also not considered the contribution of finite thickness because our calculation
of the dispersion is based on the SMA which is not the best for quantitative evaluations. 
Nonetheless, our study has merit in showing the formation of primary magnetorotons at very low wavevector with very low energies  
as the SMA has proved itself a good approximation in the case of the primary sequence of states at $\nu = n/(2n+1)$: 
It predicts correct numbers \cite{Park00,Scarola00} of magnetorotons,  
shows correct qualitative behavior at low momenta, and provides comparable (within $10$--$35\,\%$ deviation) estimation of energy
of the primary magnetoroton to the same predicted in a more robust excitonic theory of the CFs.
The determination of collective modes using standard \cite{Dev92,Scarola00,Dwipesh09} inter-$\Lambda$L excitons of CFs will not provide the lowest 
energy collective mode for the states studied
here, as it presumably involves intra-$\Lambda$L excitations.

The resonant inelastic light scattering experiments (RILSE) \cite{Dujovne03,Gallais06} using depolarized geometry where the directions of polarization of the
incident and scattered light are perpendicular to each other find very low energy (below Zeeman energy) modes in the filling factor range
$2/5 >\nu >2/7$. Not only the spin excitations but also the spinless 
excitations are selected in the depolarized spectra. On the other hand, polarized spectra in which the polarizations of the incident and scattered light
are parallel selects only the spinless excitations. Therefore observing very low energy modes in the polarized spectra of the RILSE at the filling factors
such as $\nu = 4/11$, $4/13$, and $5/13$ will prove the presence of the low energy magnetorotons in the collective
modes of excitations for these states. Surface acoustic wave experiments \cite{Kukushkin09} which are performed in determining the dispersion of the 
collective modes at $\nu =2/5$, 3/7, and 4/9 will
be very much suitable in determining the low energy magnetorotons predicted here. These excitations should also be accessible in time domain 
capacitance spectroscopy \cite{Dial10}.


In conclusion, we show that the magnetorotons form at very low energies in the dispersion of the spinless neutral collective modes at $\nu = 4/11$, 4/13, 5/13, 5/17, and 3/8, 
described by the FQHE states with unconventional topology. The observation of such low energy modes will be in support with
the presence of unconventional FQHE states between
two conventional FQHE states in the lowest Landau level.

We are grateful to J. K. Jain for his useful comments.
We acknowledge financial support from CSIR, Government of India.

\end{document}